\documentclass[journal]{IEEEtran}
\usepackage{tabularx,booktabs}
\usepackage{graphicx}
\usepackage{float}
\usepackage{stfloats}
\usepackage{subfig}
\usepackage{amsmath}
\usepackage{amsmath,amssymb}
\usepackage{makecell}
\usepackage{dsfont}
\usepackage{mathrsfs}
\usepackage{breqn}
\usepackage{enumerate}
\usepackage{circuitikz}
\usepackage{algorithm}
\usepackage{algpseudocode}
\usepackage{cite}

\begin{document}

\title{Load Oscillating Attacks of Smart Grids:~Demand Strategies and Vulnerability Analysis }
\author{Falah~Alanazi,~\IEEEmembership{Student~Member,~IEEE,}
        Jinsub~Kim,~\IEEEmembership{Member,~IEEE,}
        and~Eduardo~Cotilla-Sanchez,~\IEEEmembership{Senior~Member,~IEEE,}%
\thanks{The authors are with the School of Electrical Engineering and Computer Science, Oregon State University, Corvallis,
OR, 97331-5501 USA e-mail:~\{alanazfa,kimjinsu,ecs\}@oregonstate.edu.}%
}
\maketitle

\begin{abstract} We investigate the vulnerability of a power transmission grid to load oscillation attacks. We demonstrate that an adversary with a relatively small amount of resources can launch a successful load oscillation attack to destabilize the grid. The adversary is assumed to be able to compromise smart meters at a subset of load buses and control their switches. In the studied attack scenarios the adversary estimates the line flow sensitivity factors (LFSFs) associated with the monitored tie lines by perturbing a small amount of load at compromised buses and observing the monitored lines flow changes. The learned LFSF values are used for selecting a target line and optimizing the oscillation attack to cause the target line to trip while minimizing the magnitude of load oscillation. We evaluated the attack impact using the COSMIC time-domain simulator with two test cases, the IEEE RTS 96 and Polish 2383-Bus Systems. The proposed attack strategy succeeded in causing 33\% of load to be shed while oscillating only 7\% of load in the IEEE RTS 96 test system, and full blackout after oscillating only 3\% of the load in the Polish test system, which is much smaller than oscillation magnitudes used by other benchmarks.

\end{abstract}


\section{Introduction}

\IEEEPARstart{S}{mart} grids were introduced to make the aging electricity infrastructure more efficient, reliable, and clean. In recent decades, policy makers and governmental initiatives spurred smart grid investments. Upgrading the aging electricity grid requires installation of new technologies at all power system levels. Modernization upgrades to the distribution system level include:~communication systems, distribution automation, local control and protection systems, and advanced metering infrastructure (AMI) \cite{EPRI}. With such improvements, the potential to dynamically control and protect distributed networks becomes more feasible. Unfortunately, these improvements introduce new dimensions to attack the power grid. One of these possible attacks is hacking the AMI to control the smart meter switches by altering the data or inserting false control data. With increased smart meter deployment at the distribution level, it becomes more likely to have cybersecurity breaches \cite{SM1,vulnerability,SM2}. Successful cyber-physical attacks to the grid that have been reported, include Stuxnet \cite{stuxnet} and the attack to the Ukraine power system \cite{Ukrain}. This paper focuses on one class of the cyber-physical attacks, i.e., load oscillation attack or switching attack, which aims to disrupt the power system operation and cause instability by compromising smart meters and exploiting their switching capability.

\subsection{Related Work}

Cyber-physical attacks are also studied in several other works \cite{Anu,carter,6089026}. Work in \cite{Anu} studies the impact of load oscillating attacks on power grid stability at the transmission level. The goal of this work is to estimate the amount of load required for the attacker to cause instability at the transmission level of the power grid, and to estimate this load fraction across multiple loading scenarios and attack characteristics. However, in this work the adversary is assumed to be capable of controlling all the loads, the attack is not systematic in its approach.
Authors in \cite{carter} investigate the impact of a microgrid load oscillating attack on the main grid stability. There, more than one scenario is tested and these scenarios cover different load fractions of the compromised microgrid loads (30\%, 50\%, and 80\%) and different attack frequencies (0.05 Hz, 0.1 Hz, 0.5 Hz, and 1 Hz). However, these scenarios choose large load fractions to oscillate because there's no optimization of magnitudes 
of the load oscillation attack.
The authors in \cite{6089026} study the effectiveness of sliding-mode based switching attacks on the stability of power systems. The proposed attack causes instability in the power system by creating a sliding mode using load switching. The authors model the power system as a small single machine infinite bus (SMIB) system and the compromised breaker as a load breaker. The success of this attack assumes full knowledge of the target generator state and a very nominal scenario where the SMIB approximation is valid. 

Other load oscillation attacks are presented in \cite{interarea} and  \cite{ESS}.
The authors in \cite{interarea} successfully show that using load oscillation attack can significantly harm the power system stability. The authors use the load oscillation attack to drive a subset of system generators out of step when inter-area oscillating modes exist in the power system. The inter-area oscillation modes excited by the initial attack are estimated by the proposed attack and based on that the adversary chooses the attack frequency.  However, the initial attack proposed in \cite{interarea} imposes a 3-phase fault at specific load bus and that could be a noticeable attack for the protection relays.
Authors in \cite{ESS} investigate a switching attack using a fast-acting energy storage system (ESS) in order to disrupt part of the power system grid. In order to design the switching signal that controls the ESS circuit breaker, this work assumes that the adversary has access to the system state variables, and knowledge of the system model under different circuit breaker states.  

\subsection{Summary of Contributions}
The goal of this work is to show that an adversary with limited capability and limited knowledge is able to cause the power system to lose a considerable amount of loads and to identify the conditions under which the system is most vulnerable to such an attack.
In order to address the limits of existing attack strategies, we propose a new class of load oscillating attacks. The proposed attack is divided into three stages: reconnaissance, design, and execution. In the reconnaissance stage, an adversary estimates the LFSF associated with the monitored tie lines by slightly perturbing the load at each compromised bus and observing the resulting change in the tie line flow. During the design stage, the adversary uses the estimated LFSF to determine the target line and to identify optimal magnitudes of load oscillation that can maximize the probability of the target line tripping. During the execution stage, the adversary switches the compromised load according to the designed optimal attack in order to generate a cascading impact of the target line tripping that will be further aggravated by the continued load oscillation. 
In contrast to existing works, the proposed approach neither requires a noticeable initiating attack to excite inter-area oscillation \cite{interarea} nor necessitates access to real time state and topology information \cite{ESS}. The proposed approach does not require full dynamic state knowledge; it only uses partial state information, which is the power flow measurements of the tie lines connected to the compromised area. 
In contrast to the load oscillation attacks in \cite{Anu,carter}, which apply uniform oscillation magnitude and frequency to all compromised buses, the parameters of the proposed oscillation attacks are optimized for triggering a target line tripping event and aggravating system stability with continued oscillation. To summarize, our main contributions are as follows. We:
\begin{itemize}
  \item  propose a load oscillation attack model and methodology that requires only 1) the ability to observe the power flow measurements of a subset of tie lines that are connected to the compromised area and 2) the ability to compromise smart meters at a subset of load buses.
  \item  formulate and evaluate a load oscillation attack as an optimization problem, demonstrating the ability to identify the required amount of load oscillation for successful attacks.
  \item  evaluate the power system's susceptibility to the proposed attack using the COSMIC simulator on both the RTS-96 and the larger scale Polish test systems with the integration of the emergency control action; we provide insight into the behavior of a nonlinear power system to these types of attacks. 
  \item  show that these systems are more susceptible to specific load oscillating frequencies. 
  \item demonstrate that with a proper optimization of attack parameters, the proposed attack caused a full blackout in the Polish test system by oscillating only 3\% of the loads compared to 25\% and 9\% for the attacks in  \cite{Anu, carter} respectively.
\end{itemize} 

The rest of the paper is organized as follows:~Section 2 presents the setup and the necessary background of the load oscillation attack. In Section 3, we present the approach of the proposed load oscillation attack. In Section 4 we present the simulation setup and results analysis. Section 5 provides the conclusions. 

\section{Background and Problem Setup}  

Throughout the paper, we denote vectors by boldface lowercase letters (e.g., \textbf{x}), matrices by boldface uppercase letters (e.g., \textbf{X}) and sets by script letters (e.g., $\mathcal{X}$). In addition, $\textbf{x}_i$ denotes the $i$-th entry of vector \textbf{x}.   

The power network can be represented as a graph   $\mathcal{H}=(\mathcal{B},\mathcal{L})$ where $\mathcal{B}$ is the set of buses and $\mathcal{L}$ is the set of lines (or transformers), and each element of $\mathcal{L}$ connects two buses. We denote the set of monitored tie lines that are connected to the compromised area by $ \mathcal{\overline{L}}$, where $ \mathcal{\overline{L}}  \subset \mathcal{L}$.  We use $\mathbf{f}_{l}$ to denote the line flow on line $l$.
Each bus $\mathbf{b}_i$ is assigned a value $\mathbf{p}_i$ that represents the active injected power. The power injection can be defined by
\begin{equation}
\label{Pinj}
\mathbf{p}_i = \mathbf{g}_i -\mathbf{d}_i
\end{equation}
where $\mathbf{g}_i$ and $\mathbf{d}_i$ are respectively the total active power generation and active power demand at bus $i$. Let $\mathcal{G}$ and $\mathcal{D}$ denote the sets of the generators and load buses, respectively. The compromised load buses are a small set of buses $\mathcal{\overline{D}}$, where $\mathcal{\overline{D}} \subset \mathcal{D}$.
      

\subsection{Adversary Model}

In this work, the adversary's main goal is to design an attack that causes the power system to lose (or shed) a significant fraction of load. We assume that the adversary compromises smart meters at a subset of load buses, which we denote by $\mathcal{\overline{D}}$, and thus is capable of shedding up to a fraction $\gamma$ of the load at each bus in $\mathcal{\overline{D}}$. Furthermore, we assume that the adversary has access to the real-time measurements of a small subset of tie lines that connect the compromised area with one of the nearest areas. We denote the set of these monitored tie lines by $ \mathcal{\overline{L}}$. We assume that the adversary monitors the flow in these lines for a while and then estimates their thermal limits. Moreover, the adversary selects a certain target line, which we denote by $l^*\in  \mathcal{\overline{L}}$, such that tripping of the $l^*$ can potentially trigger cascading impacts.  

Recent cyber-physical attacks on power grids demonstrate that adversaries can in practice gain the aforementioned capabilities. Examples include 
the Ukrainian power grid  \cite{Ukrain}. 
The research group in \cite{ami} shows that smart meters could be hacked if the adversary had the smart meter ID, password, and knowledge of the communication protocol and software programming. Smart meters could be hacked as easily as mimicking communication devices to learn how to communicate with the smart meter. Malware could be spread to other smart meters through compromised smart meters to allow easier access \cite{meter}. The adversary could determine appropriate line measurement by direct intrusion or estimating information through accessing associated sensor devices or communication links \cite{attackTx,interarea}.

\begin{algorithm}[!t]
\caption{Load Oscillation Attack}
\begin{algorithmic}[1]
\State initialize $f_{os}$, $\mathcal{\overline{L}}$ and $\mathcal{\overline{D}}$
\For  {$i \in  \mathcal{\overline{D}}$}
\State $ \Delta \mathbf{d}_i =  \mathbf{d}_i + \zeta$
\State $\hat{a}_{li} =  \frac{\Delta \mathbf{f}_l}{\Delta \mathbf{d}_i}  $ for all $l\in \mathcal{\overline{L}}$
\EndFor
\State Solve the optimization problems (\ref{line_select}) and (\ref{opti_line_select}) to determine the target line   
\State Load oscillation attack starts at $t = 0$
\While { t $<~ t_{max}$}
\State Solve the optimization problem (\ref{E8}) at $t = t_{start}$
\If {$t_{start}$ $\leq$ t $\leq$ $t_{end}$}
\State s(t) = 0.5[sign(sin($\pi f_{os} $t))+1]
\If { s(t) = 1 }
\State Switch off the compromised load
\ElsIf{ s(t) = 0 }
\State Bring the compromised load into service
\EndIf
\EndIf
\State t = t + 1/$f_{os}$
\EndWhile
\end{algorithmic}
\end{algorithm}

\section{Load Oscillation Attack} 

The proposed attack consists of three stages: the reconnaissance stage, in which the LFSF is estimated; the design stage, in which the target line and the attack frequency and magnitude are determined; the execution stage, in which the oscillation attack is started.

\subsection{Reconnaissance Stage}
In the reconnaissance phase the adversary aims, without being detected, to estimate the LFSF. The LFSF measures how a change flow at line $l$ is affected by a change in each compromised load. The adversary starts the attack by identifying the compromised loads and monitored tie lines and then calculating the LFSF: 
\begin{equation}
\label{a_est}
\hat{a}_{li} =  \frac{\Delta \mathbf{f}_l}{\Delta \mathbf{d}_i} 
\end{equation}
for every $i\in \mathcal{\overline{D}}$ and $l\in \mathcal{\overline{L}} $, where 
$ \Delta \mathbf{d}_i$ is the change in the compromised load $i$ and $\Delta \mathbf{f}_l$ is the change on line $l$. The adversary does this by perturbing a very small portion $\zeta$ of $ \mathbf{d}_i$ and observing $ \Delta \mathbf{f}_l$. This means that the adversary can use (\ref{a_est}) to compute $ a_{li}$ for all compromised loads and tie lines. This perturbation is difficult to distinguish from regular load variation due to its small magnitude. 

In addition, the adversary estimates the monitored tie lines' thermal limits $\mathbf{\hat{f}}_{l}^{max}$ by observing the flow in these lines for a long period. 
Steps 1-5 in Algorithm 1 represent the reconnaissance phase.

\subsection{Design Stage}
In the design stage, the adversary uses the LFSF and $\mathbf{\hat{f}}_{l}^{max}$ calculated in the reconnaissance stage for guidance in identifying a target line that the adversary can possibly cause to trip using the load oscillation attack. In addition, given a constraint on load oscillation capability of the adversary, the adversary determines the optimal load oscillation magnitudes for triggering tripping of the target line. First, the adversary aims to predict the maximum power flow that could be generated in each of the monitored lines. The adversary uses the compromised loads and the estimated LFSF to formulate an optimization problem with the objective of maximizing $ \mathbf{f}_l$. It is important to note that the adversary is solving the optimization problem for each monitored line $ \mathbf{f}_l$. The optimization formulation is then described as follows:

\begin{equation}
\label{line_select}
\begin{aligned}
\max_{\Delta \mathbf{d}_i, \mathbf{f}_l}  \mathbf{f}_l\\
\textrm{s.t.} \quad & \gamma_i  \mathbf{d}_i \leq \Delta \mathbf{d}_i \leq 0,~~ \forall i \in \mathcal{\overline{D}}\\
\quad & \mathbf{f}_l = \mathbf{f}_l^0+\sum_{i\in \mathcal{\overline{D}}}{a_{li} \Delta \mathbf{d}_i}\\
\end{aligned}
\end{equation}
where $\gamma_i$ is the proportion of loads that are compromised. Thereafter, the adversary uses the solution from (\ref{line_select}) for all monitored lines to search for the line that is most likely to violate its thermal limits. 
In this case, the adversary selects the target line as follows:
\begin{equation}
\label{opti_line_select}
\begin{aligned}
\\
l^*= \operatorname*{arg\,max}_{l \in \mathcal{\overline{L}}} ~\frac{\mathbf{f}_l^{*}}{\mathbf{\hat{f}}_{l}^{max} }      \\
\end{aligned}
\end{equation}
where $\mathbf{f}_l^{*}$ is the optimal solution of (\ref{line_select}) and $l^*$ is the selected target line. After selecting the target line, the adversary uses the solution of (3) for $f^*_{l^*}$  to determine the magnitude of load oscillation at each compromised load bus that maximizes the line flow in the target line. 
Note, the adversary optimizes the power flow in the transmission line in only one direction, outgoing from the compromised area, to match the direction in which excessive power flows when compromised loads are shed.

\subsection{Execution Stage}
In the execution stage the adversary follows the switching signal $s(t)$, which oscillates between 0 and 1: 
\begin{equation}
\label{switch}
s(t) = 0.5~[~\text{sign}(\text{sin}(2\pi f_{os}t))+1] 
\end{equation}
where $ f_{os} $ is the attack oscillation frequency. When the switching signal is equal to 1 the adversary opens the switches associated with the load at the compromised buses $i$ according to the attack designed by solving (\ref{line_select}) and (\ref{opti_line_select}). On the other hand, the adversary closes all the compromised switches when the switching signal is equal to 0. The adversary continues oscillating the load throughout the attack period as an effort to drive the system state away from the stable region, even after the target line trips. The feasibility of the linear optimization solution from (\ref{line_select}) and (\ref{opti_line_select}) that pushes the magnitude of the target line flow above its relay threshold depends on the system operating point. Therefore, the adversary can adjust the time of the attack such that the line flow is near its thermal limit. For example, the adversary may choose to launch the attack when the daily demand is at peak.
The effect of this attack on the power system varies depending on the adversary behavior. Therefore, we consider three different attack implementations, which represent different possible actions the adversary may use. Here we compare the impact of using one oscillation frequency versus alternating between two oscillation frequencies in the attack. We assume that an adversary may choose to use two different behaviors in one attack. For example, the adversary could start the attack with high oscillating frequency and then after a specific period the adversary changes the oscillating frequency into a low or medium frequency.

\section{Experiments} 
We evaluate both the load oscillation attack model and its impact on the IEEE RTS-96 bus system \cite{RTS96} and Polish 2383-Bus System \cite{poland}. We used the Cascading Outage Simulator with Multiprocess Integration Capabilities (COSMIC) \cite{COSMIC} to conduct the experiments. COSMIC simulates both cascading failures and the full ac power flow dynamics of the system, including physical models of generators and their control systems, emergency protection relays, and loads. 

\begin{figure}[!t]
\centering
\includegraphics[width=0.85\linewidth,height=0.20\textheight]{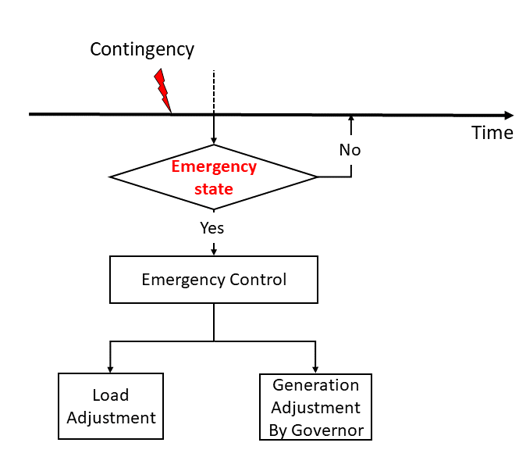}
\caption{Integrating emergency control with dynamic time domain simulation.}
\label{fig: EC}
\end{figure}

In this work we evaluate the performance of the proposed attack using the time domain simulator COSMIC, with and without emergency control integration. The emergency control is an optimal power flow using a dc power flow approximation integrated to the time domain simulator to quickly find a dispatch that could help out in simulated real time. Figure \ref{fig: EC} shows how the emergency control is integrated into the dynamic simulator. The emergency control activates only if the system falls into an emergency state (in our case any branch tripping). Our goal here is to show the difference in the power system's time domain simulator response to the attack with and without emergency control presence.

The emergency control is formulated in a traditional way to minimize the amount of load shedding and generation adjustment in a grid in response to the amount of transmission line overloads \cite{EC,EC1,EC2}. The solution of the optimization problem determines the amount of load need to be shed. It is necessary to minimize the amount of load shedding as soon as possible because the large difference between the generation and demand can increase instability. The emergency control is formulated as follows: 

\begin{align}
\min_{\Delta \mathbf{d}, \Delta \boldsymbol{ \theta }, \Delta \mathbf{g}} \quad & -1^T \Delta \mathbf{d} \label{E9}\\
\textrm{s.t.} \nonumber\\
\quad & (\mathbf{g}^0+\Delta \mathbf{g})-(\mathbf{d}^0+\Delta \mathbf{d}) = \mathbf{B}(\boldsymbol{\theta}^0+\Delta \boldsymbol{\theta}) \label{E10}\\
\quad & \mathbf{f} = (\mathbf{D}\times \mathbf{A})(\theta^0 + \Delta \theta) \label{E11}\\
\quad & \mathbf{f}^{min} \leq \mathbf{f} \leq \mathbf{f}^{max} \label{E12}\\
\quad &  \mathbf{g}^{ min} \leq \mathbf{g}^0 +\Delta \mathbf{g} \leq  \mathbf{g}^{ max} \label{E13}\\
\quad &  - \mathbf{d}^0 \leq \Delta \mathbf{d} \leq  0  \label{E14} \\
\quad &  \Delta \boldsymbol{\theta}_i = 0, \forall i \in  \mathcal{I}_{ref} \label{E15} \\ \nonumber
\end{align}
where $\Delta \mathbf{g}$, $\Delta \mathbf{d}$, $\mathbf{g}^0$, $ \mathbf{d}^0 $, and $\theta^0$ in (\ref{E10}) are column vectors of generation adjustments, load adjustments, initial generation , initial load, and initial voltage phase angle. (\ref{E11}) and (\ref{E12}) show the line flow computations and limits. Inequality constraints (\ref{E13}) and (\ref{E14}) impose the desired limits of generation and load adjustments. (\ref{E15}) ensures that $ \Delta \boldsymbol{\theta}$ for the reference buses is zero.

The load model used in this work is the traditional ZIP model or the polynomial load model, which consists of the sum of three known load models: constant impedance (Z), constant current (I), and constant power (P). The ZIP model expressions for both active and reactive power are:

\begin{equation}
\label{E16}
\mathbf{p_i}=\mathbf{p}^{r}_i \bigg[\mathbf{c}_p^z \Big( \frac{\mathbf{v}_i}{ \mathbf{v}^{r}_i}\Big)^2+\mathbf{c}_p^i\Big( \frac{\mathbf{v}_i}{ \mathbf{v}^{r}_i}\Big)+\mathbf{c}_p^p \bigg]
\end{equation}
\begin{equation}
\label{E17}
\mathbf{q}_i=\mathbf{q}^{r}_i \bigg[\mathbf{c}_q^z \Big( \frac{\mathbf{v}_i}{ \mathbf{v}^{r}_i}\Big)^2+\mathbf{c}_q^i\Big( \frac{\mathbf{v}_i}{ \mathbf{v}^{r}_i}\Big)+\mathbf{c}_q^p \bigg]
\end{equation}          
where $\mathbf{c}_p^z $, $\mathbf{c}_p^i $, and $\mathbf{c}_p^p$, are the ZIP model coefficients for the active power, and $\mathbf{c}_q^z $, $\mathbf{c}_q^i $, and $\mathbf{c}_q^p $  are the ZIP model coefficients for the reactive power of the ZIP model;  $\mathbf{p}^{r}_i $ and $\mathbf{q}^{r}_i $ are the active and reactive powers at rated voltage $\mathbf{v}^{r}_i $; $\mathbf{p}_i $ and $\mathbf{q}_i $ are the active and reactive powers at operating voltage $\mathbf{v}_i $. We choose to use the ZIP model coefficients as illustrated by 
\cite{ZIP_COFF} because these coefficients describe the behavior of the modern loads during the steady-state and varying voltage conditions accurately. 

\subsection{Validating Simulator Settings in Comparison to Benchmarks}
In this section, we compare the COSMIC simulator performance with commonly used cascade failure models in the literature as well as historical data from the real world. The goal of this comparison is to validate the settings of COSMIC in this experiment in comparison to benchmarks. The benchmarking is based on the probability distribution of demand loss. 
The benchmark using a subset of data from \cite{benchmark}, is comprised of PCM \cite{PCM}, Manchester Model\cite{manch}, PRACTICE \cite{pract}, TransCARE model \cite{Trans}, and historical data \cite{historical}. All cascading failure simulations in the benchmark are performed on the RTS 96 bus system.
To compare the COSMIC simulator as used here with the benchmark,  1200 N-2 contingency scenarios were simulated based on the same  methodology and settings used in \cite{benchmark}. All simulations start from the same system state. 

\begin{figure}[!t]
\centering
\includegraphics[width=0.52\textwidth]{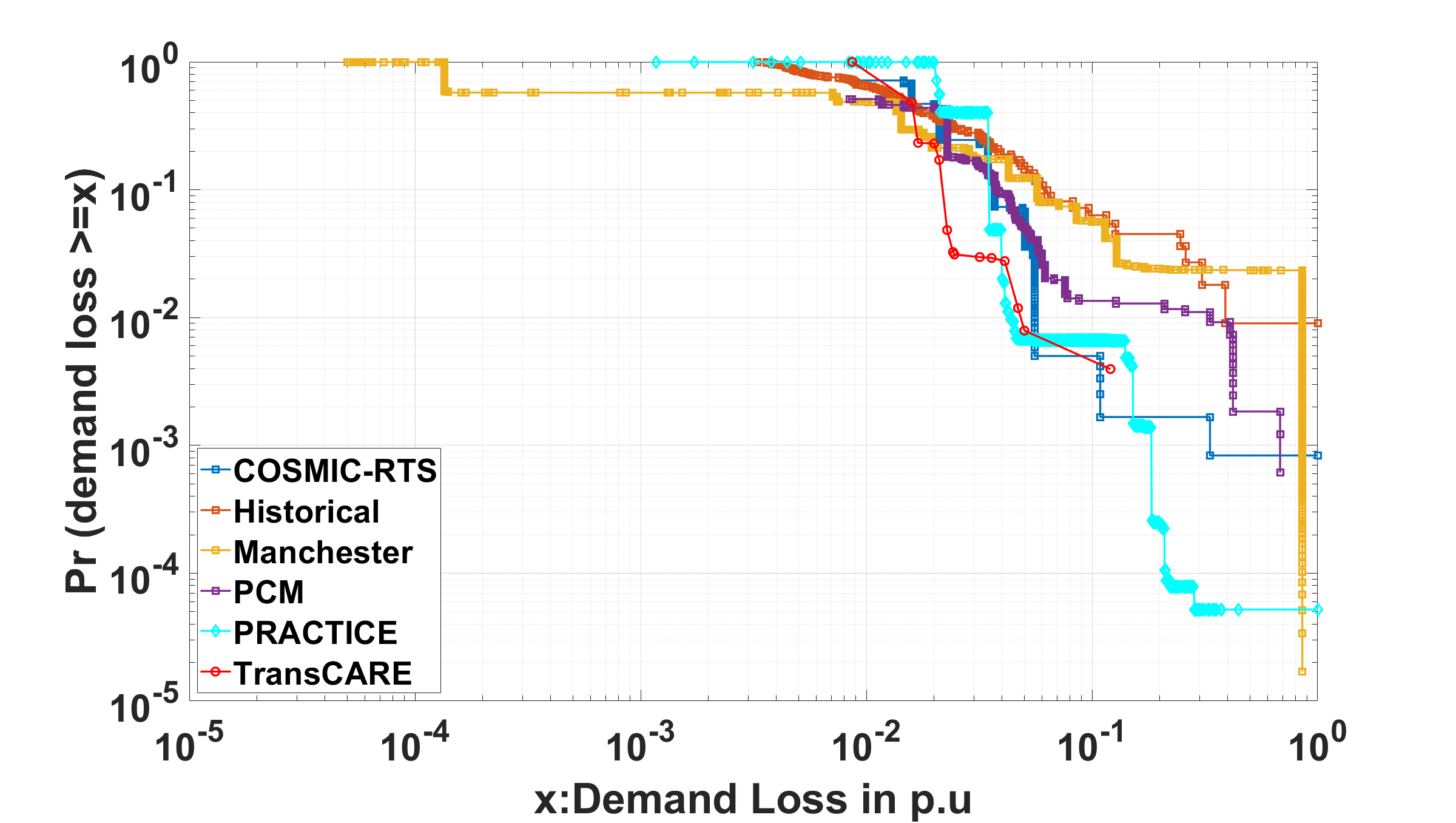}
\caption{Distribution of load shed for different models, adapted with data from \cite{benchmark}.}
\label{fig: Distribution of load shed-orig}
\end{figure}

Figure \ref{fig: Distribution of load shed-orig} shows the distribution of demand loss in the form of the survival function of data from COSMIC and the selected methods. The survival function is a complementary cumulative distribution function of the demand loss and this distribution is conditioned on the demand loss being greater than a given value.  From Figure \ref{fig: Distribution of load shed-orig} we can compare the predicted frequencies of all blackout sizes for different simulators. This allows us to compare and validate the overall setup of COSMIC with commonly used cascade failure models. The probability is equal to 1 for all methodologies when demand loss approaches 0. The log-log plot is used to show the large blackouts with smaller probabilities. We can see in Figure \ref{fig: Distribution of load shed-orig} that as the blackout size increases, the survive function decreases, which indicates that the frequency of blackouts decreases as their size increases. In addition, we can observe that for small blackouts, all the methodologies have similar result. Moreover, Figure \ref{fig: Distribution of load shed-orig} shows that the probability of having medium and large blackouts have more of a disagreement for each cascade failure model, but COSMIC, as tuned for these experiments, largely follows the benchmark trends.

\subsection{Attack Classes}
\begin{figure}[t]
\centering
\includegraphics[width=0.75\linewidth,height=0.32\textheight]{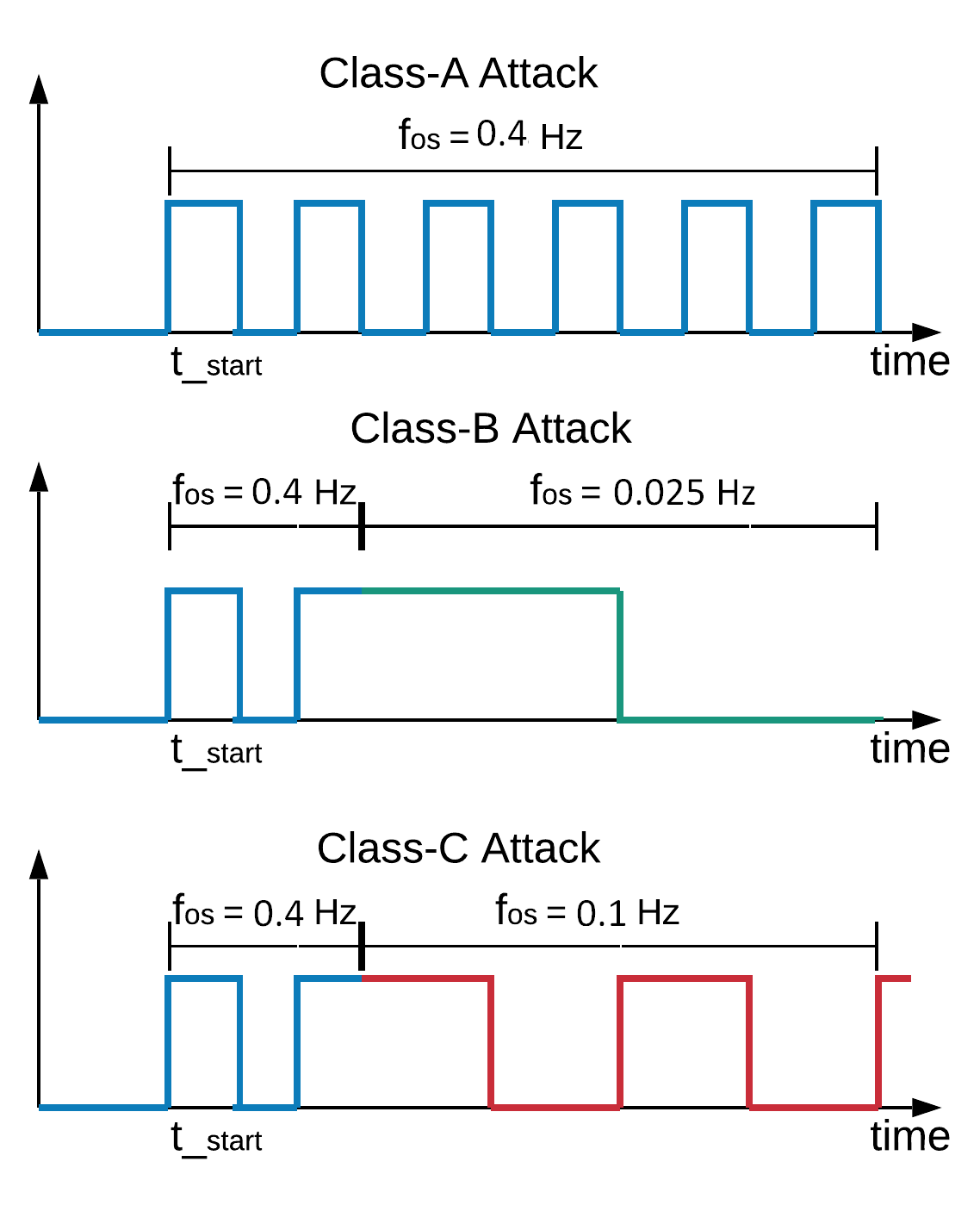}
\caption{Load oscillation attack classes.}
\label{fig: Attack2}
\end{figure}

In this paper, we implement three different attack scenarios. These scenarios study the impact of using one oscillation frequency versus alternating between two oscillation frequencies in the attack. As illustrative example we divide the attacks into three classes, and we select the oscillation frequencies based on the analysis done in \cite{Anu,carter} as shown in  Figure \ref{fig: Attack2}:

\begin{itemize}
  \item \textbf{Class A} using a fixed high oscillating frequency at $f_{os}=0.4~ Hz$. 
  \item \textbf{Class B} using two oscillating frequencies: high $f_{os}=0.4~ Hz$ and low $f_{os}=0.025 ~Hz$. 
  \item \textbf{Class C} using two oscillating frequencies: high $f_{os}=0.4~ Hz$ and medium $f_{os}=0.1 ~Hz$.
\end{itemize}
In addition, later in this section we study the impact of using different oscillating frequencies ranging from 0.1 to 0.8 Hz in the proposed attack classes. We select this frequency range because the interconnected power system exhibits inter-area oscillation modes ranging from 0.1 to 0.8 Hz during disturbances \cite{interarea,OSC2,OSC}.

As we can see in Figure \ref{fig: Attack2}, the power system begins operating at steady state. The adversary starts the attack at $t_{start} = 10$ sec and ends at $t_{end} = 70$ sec. We assume that the adversary can control up to 8\% of the system loads and no more than 25\% of each of the compromised loads ($ \gamma = 0.25$). Controlling 8\% or more of the system load can be plausibly achieved by reaching only three closely connected load buses \textemdash such as 318, 316, and 314 \textemdash in the RTS 96 test case. Classes B and C start with high oscillation frequency for the first 10 seconds, then switch to low or medium oscillation frequency. During the attack some of the protective relays may shed either compromised or uncompromised loads and that will reduce the total load of the grid. 


\subsection{IEEE RTS-96 Bus System}
The proposed attack is first performed on the RTS-96 bus system. 
The RTS-96 case 
consists of 73 buses and is divided into three almost identical areas: 1, 2, and 3.
The RTS-96 case is verified to be N-1 secure. For simulation purposes we consider the smart meter switches in Area 3 to be vulnerable to the load oscillating attack and the tie lines that connect the compromised area with Area 1 and 2 to be the monitored lines. 

All attack experiments last for 130 seconds divided as follows: the first 10 seconds is normal operation followed by 60 seconds of oscillating attack, and then the simulation termination will be 60 seconds after the attack ends. It is important to note that the simulation time step-size is not fixed. COSMIC selects a small time step-size during transition periods and a larger step-size when the system response trends to a steady state \cite{COSMIC}.

\begin{table}[t]
\begin{center}
\caption{An example target line estimated thermal limit, its initial power flow before the attack, and its predicted power flow after the attack.}
\label{tab:table1}
 \begin{tabular}{ l c  c  c  c} 
 \toprule
 \multicolumn{1}{>{\centering\arraybackslash}m{14mm}}{\textbf{\bfseries Tie Line Number}} & 
  \multicolumn{1}{>{\centering\arraybackslash}m{14mm}}{\textbf{$\mathbf{\hat{f}}_{l}^{max} {(MW)} $}} & 
  \multicolumn{1}{>{\centering\arraybackslash}m{13mm}}{\textbf{$\mathbf{f}_l^0~\mathbf{(MW)} $}} & 
  \multicolumn{1}{>{\centering\arraybackslash}m{13mm}}{\textbf{$\mathbf{f}_l^{*}$}} &   
  \multicolumn{1}{>{\centering\arraybackslash}m{13mm}}{\textbf{$\mathbf{f}_l^{*}$/ $\mathbf{\hat{f}}_{l}^{max}$}}   \\
 \midrule
 118 & 500 & -38  & 211 & 0.42 \\ 
 119 & 500 & 103.3  & 357.5 & 0.71\\
\bottomrule
\end{tabular}
\end{center}
\end{table}

\subsubsection{Attack Performance without Emergency Control}
Following Algorithm 1 Steps 1 to 5 (reconnaissance phase), as presented in Section III, the LFSFs are estimated. Using the estimated LFSFs and $\gamma = 0.25$, the maximum power flow that could be generated in each of the monitored lines $\mathbf{f}_l^{*}$ is predicted by solving (\ref{line_select}), as we can see in Table \ref{tab:table1}. The result shows that the adversary is capable of increasing the power flow in either tie line, however, solving (\ref{opti_line_select}) shows that the tie line 119 is the most likely to violate its thermal limits. The tie line 119 is selected as the target line. After selecting the target line, we determine the magnitude of load oscillation at each compromised load by using the solution of (\ref{line_select}) for $l^*$.

\begin{figure*}[t]
\centering
\includegraphics[width=17.5cm, height=9cm]{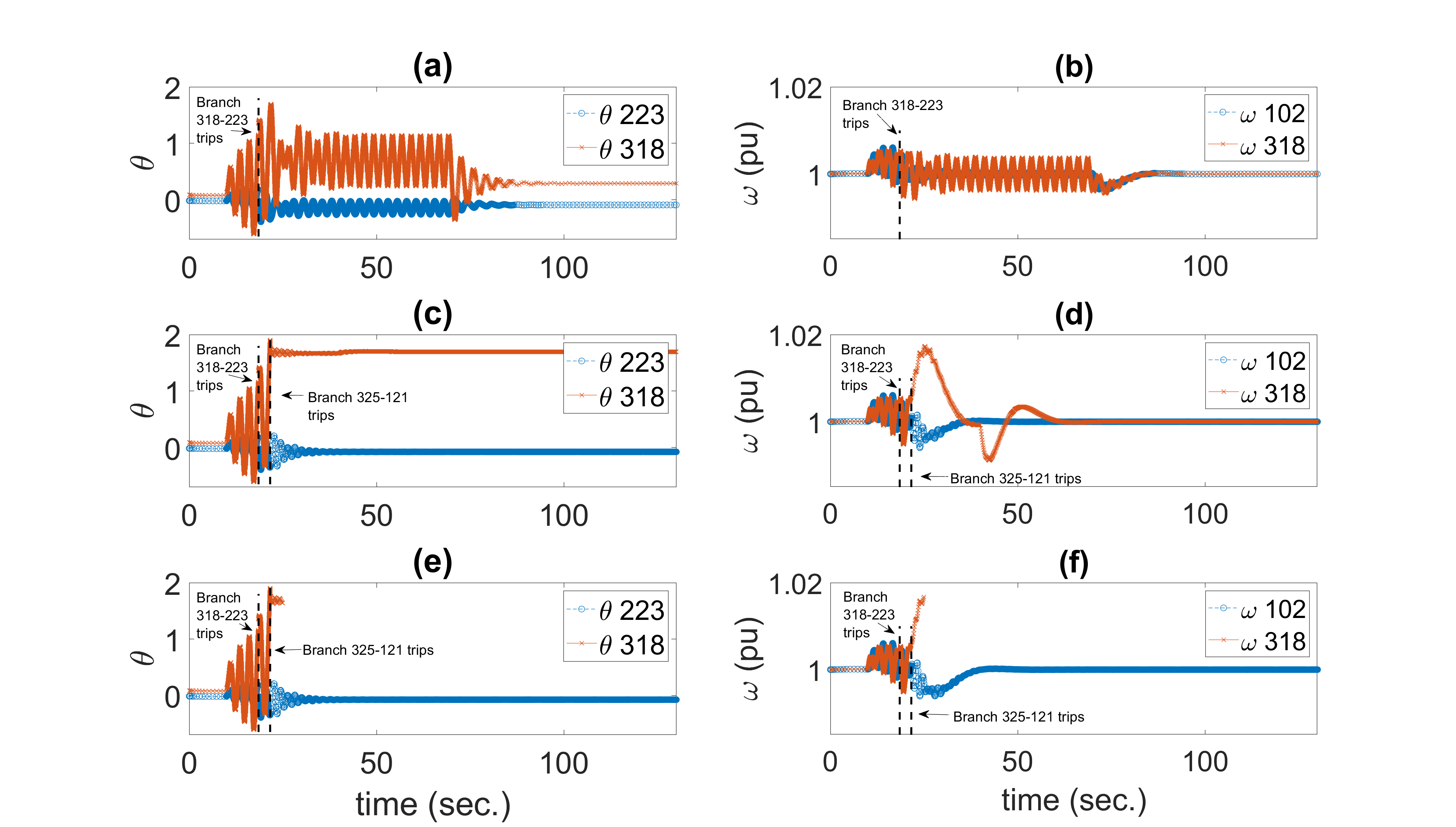}
\caption{Impact of load oscillation attack Class A (a \& b), Class B (c \& d), and Class C (e \& f) on voltage angle for Buses 223 and 318, and angular speed for Generators 102 and 318.}
\label{fig: omega_all_classes_NO_EC3 }
\end{figure*}

The 3 attack classes presented in the previous subsection are tested with $\gamma = 0.25$, which means that the adversary is able to control up to 25\% of each of the compromised loads, which are in turn 8\% of the system load. The impacts of these attacks on the voltage phase angle of both ends of the target line $ \theta_{223}$ and $\theta_{318} $ are shown in Figure \ref{fig: omega_all_classes_NO_EC3 }(a), (c), and (e). Figure \ref{fig: omega_all_classes_NO_EC3 }(b), (d), and (f) demonstrate the impact of the oscillating attack on the angular speed $\omega $ of two generators, one within the compromised area (Gen 318) and another one far from it (Gen 102).

\subsubsection{Class A Attack} 

Figure \ref{fig: omega_all_classes_NO_EC3 }(a) shows the voltage phase angle of both ends of the target line $ \theta_{223}$ and $\theta_{318} $ during the Class A attack. 
The attack causes both $ \theta_{223}$ and $\theta_{318} $ to oscillate together; however, the oscillation magnitude in $\theta_{318}$ in the positive side increases with every cycle. This indicates that the power flow from Area 3 to Area 2 increases with every cycle. When the target line trips at $t=18.5$ sec, $ \theta_{223}$ and $\theta_{318} $ lose their relationship and start to oscillate separately in different manners. The dashed lines and the arrows in Figure \ref{fig: omega_all_classes_NO_EC3 } identify the location of the tripping events in the result. Figure \ref{fig: omega_all_classes_NO_EC3 }(b) demonstrates the impact of the Class A attack on $ \omega_{102}$ and $\omega_{318}$. As shown, $ \omega_{102}$ and $\omega_{318}$ are stable until the attack is initiated and then they both start oscillating. At t = 18.5 sec the target line trips and that significantly reduces the $ \omega_{102}$ oscillation. For $\omega_{318}$ we observe an increase in the oscillation after the line trips for two cycles but after that the oscillation decreases.

\subsubsection{Class B Attack} 

This class investigates the impact of using two oscillating frequencies: fast (0.4 Hz) and slow (0.025 Hz). Similar to Class A, the grid begins with normal operation at the precomputed steady state. The attack is initiated at $t = 10$ sec. but the adversary changes the $f_{os}$ from 0.4 Hz to 0.025 Hz at $t = 20$ sec. Figures \ref{fig: omega_all_classes_NO_EC3 }(c) and (d) show the impact of the Class B attack on $ \theta_{223}$, $\theta_{318} $, $ \omega_{102}$ and $\omega_{318}$ respectively. We can see in Figures \ref{fig: omega_all_classes_NO_EC3 }(c) and (d) that the target line trips at $t=18.5$ sec. and the tie line connecting Area 1 and 3 trips at $ t=21.5$ sec. Although this attack compromises the same loads as the previous attack, all the tie lines connected to Area 3 trip and the system splits into two islands, Area 1 and 2 as Island-1 and Area 3 as Island-2. Figure \ref{fig: omega_all_classes_NO_EC3 }(d) shows that Generator 318 in Island-2 is no longer synchronized with Generator 102 in Island-1. We notice an interesting behavior when the system splits into two islands; the response of $\omega_{318}$ deviates up significantly because the compromised loads are off when the system splits into two islands. Although Island-2 has more generation than demand and that causes a sharp increase in $\omega_{318}$, the governors and exciters are sufficient to bring $\omega_{318}$ back into 1 p.u. (60 Hz). In addition, we can see that $\omega_{318}$ decreases sharply at t = 40 sec. This is because the adversary brings more demand to the system by reconnecting the compromised loads. However, the control system succeeds to bring $\omega_{318}$ back into 1 p.u. again.

\subsubsection{Class C Attack}

We study the possible case of using two different attack frequencies, fast in the first 10 seconds and medium after that. This attack has a significant impact on the system, causing the system to lose 33\% of its load and generation. Figure \ref{fig: omega_all_classes_NO_EC3 } (e and f) shows that the power grid splits into two islands like the Class B attack. However, the governor and exciter controls associated with Area 3 generators are not able to return the compromised island into a secured state. At the beginning of the attack the adversary uses (\ref{switch}) with a switching frequency of 0.4 Hz until the target line trips. After that the adversary changes its attack frequency into a slower attack which is 0.1 Hz. As a result of this change the system splits into two islands; however, the island that contains the compromised load is lost.

\begin{figure}[t]
\includegraphics[width=0.45\textwidth]{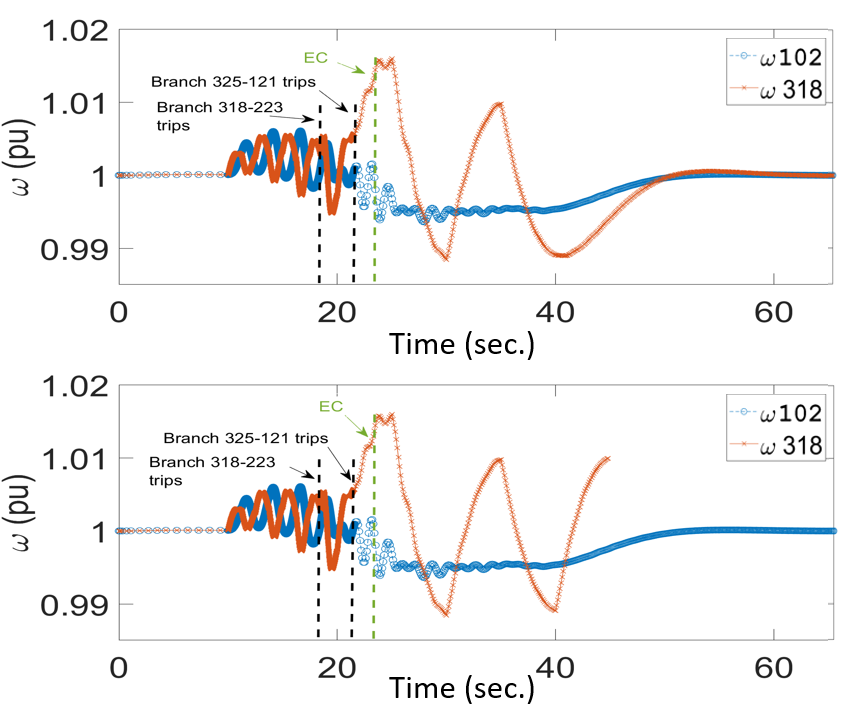}
\caption{Impact of the Class C attack for 25-second attack period (top panel) and 60-second attack period (bottom panel) on angular speed for Generators 102 and 318 with emergency control integrated.}
\label{fig: omega_all_classes_EC}
\end{figure}

\subsubsection{Attack Performance with Emergency Control}
The following discusses the impact of the proposed emergency control algorithm on mitigating the load oscillating attacks. The emergency control is assumed to operate 5 seconds after a branch trips and given that in Class A and B attacks no loads or generations is lost, we will focus on the Class C attack. Figure \ref{fig: omega_all_classes_EC} illustrates generator angular speed for Generator 318 and Generator 102 during a Class C attack for 25 seconds (top panel) and 60 seconds (bottom panel) respectively, with emergency control.

Without emergency control, both attacks caused a separation of the network. Imbalance between generation and demand caused the loss of one island, as shown in Figure \ref{fig: omega_all_classes_NO_EC3 }-e. However, the other island was able to converge to a balanced state. In contrast, during the 25-second attack, with the emergency control help, the system succeeded in avoiding further collapse, allowing both islands converge to a secure state. For the 60-second attack, the emergency control (and the governor and exciter control) tried to restore the network to a relatively healthy condition. However, the result reveals that it was not sufficient to bring the Area 3 island to a stable state.  

\begin{figure}[t]
\centering
\includegraphics[width=1\linewidth,height=0.23\textheight]{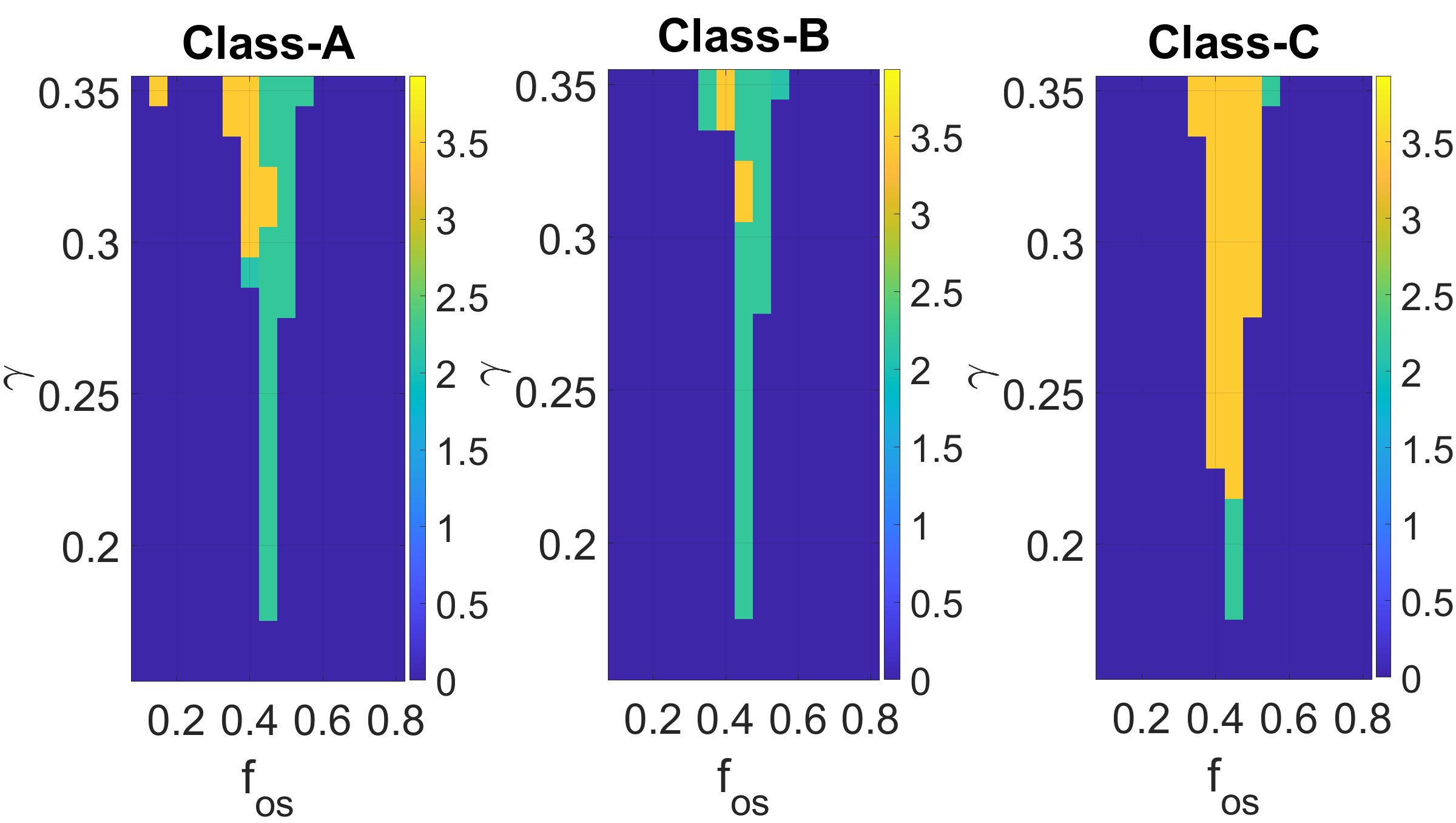}
\caption{The $\log_{10} (\text{losses})$ for load oscillation attacks with variable compromised load fractions and attack oscillation frequency.}
\label{fig: RTS_freq}
\end{figure}

\subsubsection{Sensitivity Analysis}
In this part we  study  the  sensitivity  of  the proposed  attacks  to  several  oscillating  frequencies  ranging from 0.1 to 0.8 Hz as well as different fractions of the compromised load $\gamma$. Figure \ref{fig: RTS_freq} depicts the amount of the lost load for several simulation cases for Class A, B and C, where the oscillating frequency and the compromised loads fraction are varied. The system total loads is 8550 MW. However, we use $\log_{10}$ of the lost loads to show small blackouts.
\begin{equation}
\label{log}
X = 
    \begin{cases}
      \log_{10} (\text{losses}) & \text{if} ~~ \text{ losses} > 0\\
      0 & \text{otherwise}
      
    \end{cases}
 \end{equation}
One can see is clear that the oscillating frequencies 0.4, 0.45, and 0.5 Hz cause a more significant disturbance to the power system for all proposed attacks. However, the Class C attack causes more impact on the system than Class A and B. During the Class C attack, the power system starts losing 33\% (orange color) of its total load at $\gamma$ ranging from $0.22$ and $0.28$  for frequency between $ f_{os} = 0.4$ and $0.5 $ Hz. 
The Class A and B attacks, on the other hand, are only able to cause the system to lose 33\% of its total load within a much narrower range of $\gamma$ values. We notice that both classes cause the system to lose 33\% of its total loads at $f_{os}=0.45$ Hz only at $\gamma=0.31$ and $0.32$. This is because the two tie lines that connect the compromised area with the rest of power system trip at the same time (in this situation only). At $f_{os}=0.40$ Hz the Class A attack causes the system to lose 33\% of total load when $\gamma\geq0.30$, while the Class B has the same impact when $\gamma\geq0.34$. It should be noted that the system is more susceptible to oscillating frequencies ranging between 0.4 and 0.5 Hz.

\begin{figure}[t]
\includegraphics[width=0.5\textwidth]{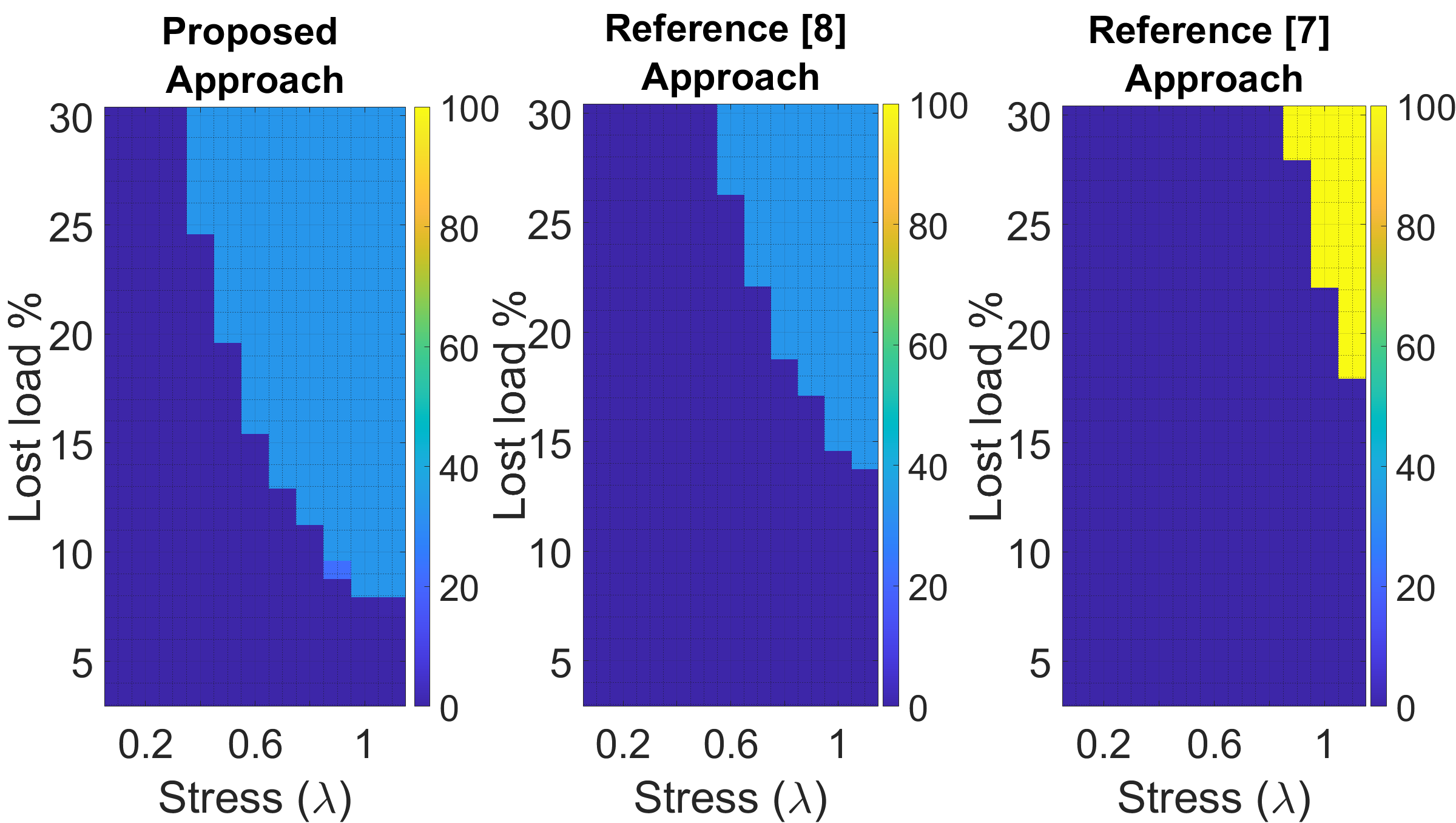}
\caption{The percentage of lost loads for load oscillation attacks with variable compromised load percentage and system stress level for the present approach compared to \cite{carter, Anu}.}
\label{fig: comp_all}
\end{figure}



\subsubsection{Validation}To illustrate the significance of our vulnerability analysis, we compare the analysis of this paper with the vulnerability analysis done by the research groups in \cite{Anu,carter}. We choose these two works because they use similar attack types and assumptions that we use in our approach. The work in \cite{Anu} chooses to distribute the adversary capability throughout all system loads, while our approach and the work in  \cite{carter} concentrate the adversary capability to a specific area in the system.

We conduct all experiments on the RTS-96 bus system and we use the COSMIC simulator to study the impact of the proposed load oscillation attacks. We compare the most effective class (Class C with $f_{os} = 0.4$ Hz and $f_{os} = 0.1$ Hz) of our proposed attacks against \cite{Anu,carter} and their most effective attacks. Since the power system resilience to external attack varies depending on the system state prior to that attack \cite{stress}, we compare the impact of all attack approaches for different system stress levels ($\lambda$) and compromised load fractions. For all simulations, we limit the adversary capability of controlling system loads to 30\% of the total system load. Similar to the attack in \cite{carter}, we choose the load buses in Area 3 to be the compromised buses for our attack.

Figure 
\ref{fig: comp_all} show the percentage of surviving loads after simulating all the load oscillation attacks proposed in this validation subsection. The dark blue color means that the system does not lose any load, the yellow color means that the system lost all the loads and the simulation did not converge. 
We can see in both figures that the system stress and the system security have an inverse relationship, as the stress increases, the amount of compromised load needed to cause the power system to lose a significant fraction of load decreases. However, we can observe from Figure \ref{fig: comp_all} that both benchmarks fail to cause any damage to the system when compromising less than 15\% of the system loads. On the other hand, we were able to cause the system to lose 33\% of its load by using less than 15\% of system loads when $\lambda$ is greater than 0.7. This observation indicates that with proper optimization of attack parameters, the load oscillation attack can disturb the power system with a limited fraction of the compromised load.
Figure \ref{fig: comp_all} shows that at high stress level $\lambda\geq 0.8$ and high compromised load percentage, the proposed attack by \cite{Anu} causes the system to become unstable and no loads survive.

When the power system is not stressed, which is $\lambda\leq 0.8$, concentrating the attack into one area of the system causes more damage as shown in Figure \ref{fig: comp_all}. The proposed work by \cite{Anu} succeeds to disturb the power system only when $\lambda\geq 0.8$.

\subsection{Polish 2383-Bus System}

To further demonstrate the capability of the proposed attack on a realistic and relatively large transmission system, we use the Polish test system \cite{poland}. The system consists of 2383 buses, 327 generators, 1826 loads, and 2896 lines. For simulation purposes we consider Area 3 to be the compromised area and the tie lines that connect Area 3 with Area 4 to be the monitored tie lines. 

\begin{figure}[t]
\includegraphics[width=0.5\textwidth]{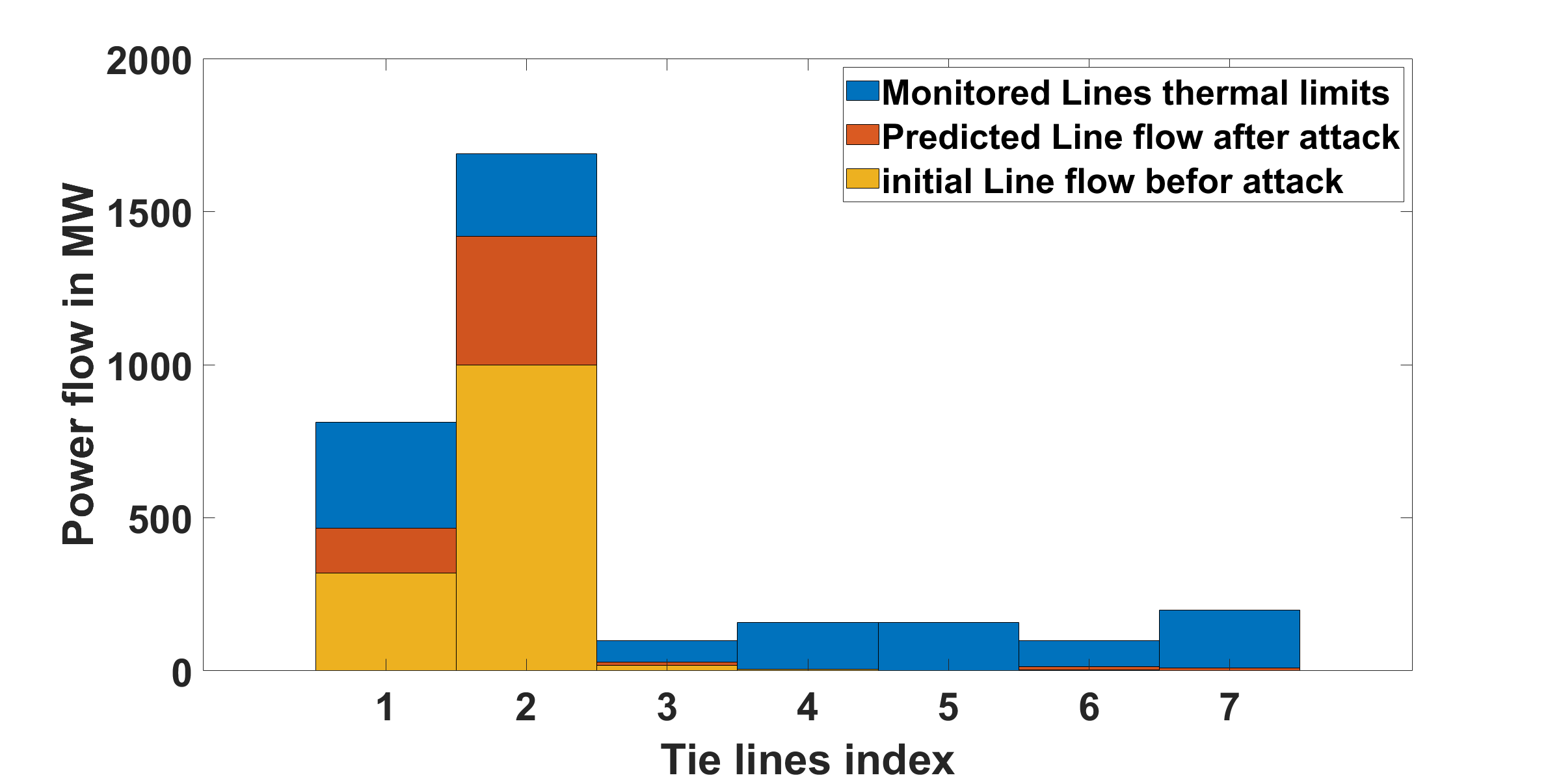}
\caption{Comparison between the initial line flow, the predicted line flow after the attack, and the estimated thermal limits for the monitored tie lines.}
\label{fig: line_s}
\end{figure} 

\subsubsection{Target Line Selection} The target line is selected following Algorithm 1 steps 1 to 6. The algorithm uses the estimated LFSF to predict the maximum power flow that can be generated in the monitored tie lines $ \mathbf{f}_l$ and then uses both $ \mathbf{f}_l$ and the estimated $\mathbf{\hat{f}}_{l}^{max}$ to select the target line. 
There are 7 lines connecting the compromised area with Area 4 and the aim is to select the line that is most likely to violate its thermal limit after the attack initiated as the target line. 
Figure \ref{fig: line_s} demonstrates the initial power flow, the predicted power flow after the attack, and the estimated thermal limits for the monitored lines. We can see in Figure \ref{fig: line_s} that most of the lines are predicted to not exceed their thermal limits. However, the algorithm selects Line 2 because it has the highest percentage  between its predicted value and estimated thermal limits. Therefore, Line 2 is selected as the target line.

\begin{figure}[t]
\centering
\includegraphics[width=1\linewidth,height=0.23\textheight]{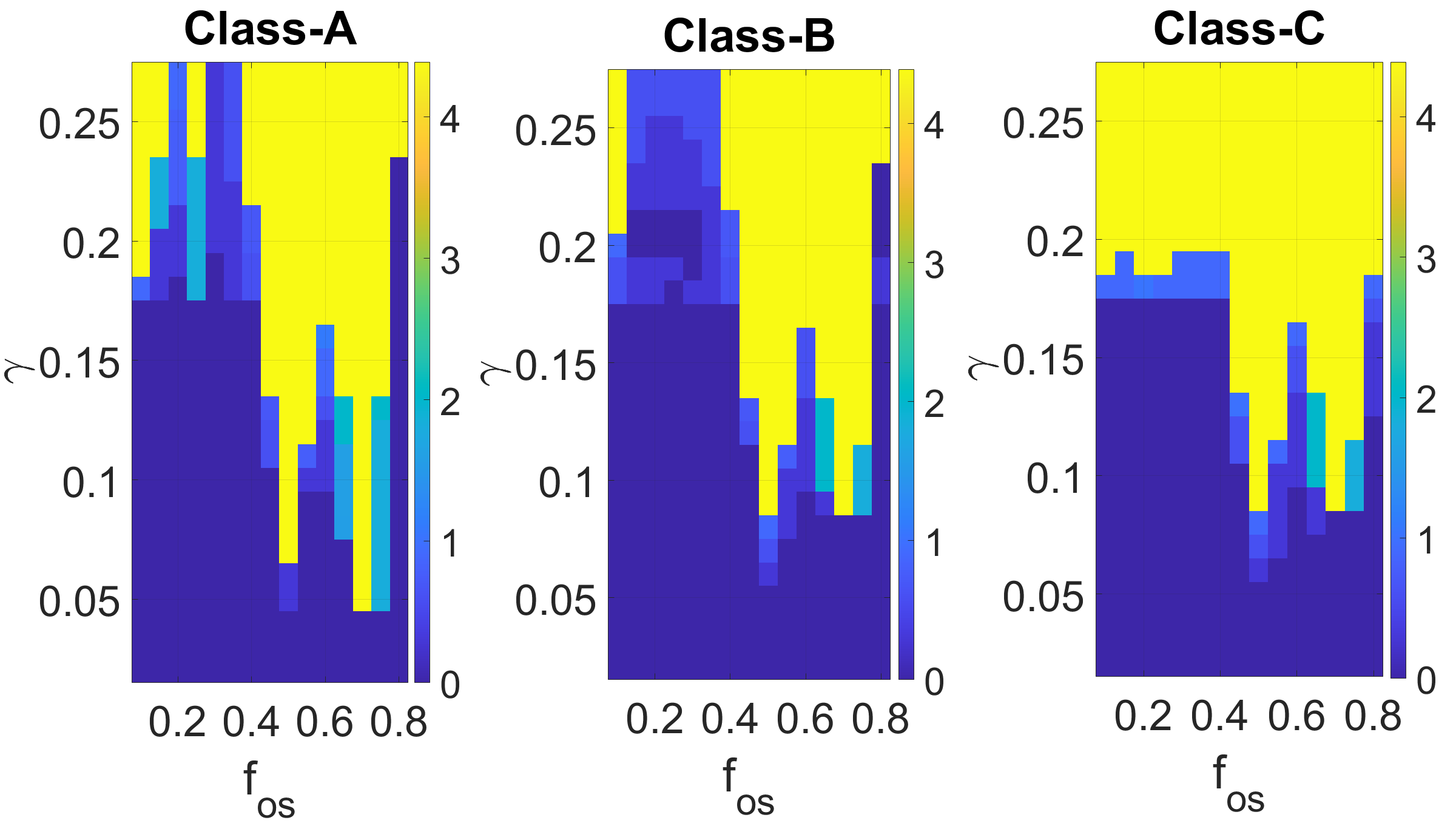}
\caption{The $\log_{10} (\text{losses})$  for load oscillation attacks with variable compromised load fractions and attack oscillation frequency.}
\label{fig: poland_freq}
\end{figure}

\subsubsection{Sensitivity Analysis} Similar to the sensitivity analysis presented in Subsection C, the impacts of using several oscillating frequencies ranging from 0.1 Hz to 0.8 Hz and several compromised load fractions $\gamma$ in the proposed attacks are studied. Figure \ref{fig: poland_freq} shows the sensitivity analysis for Class A, B, and C attacks. The system total loads is 24558 MW. We use (\ref{log}) to show the small blackouts. The proposed attack classes cause more damage on the system with high oscillating frequencies ranging from 0.45 to 0.75 Hz. It is important to observe that the proposed attacks with oscillation frequencies 0.5 and 0.7 Hz succeed in destabilizing the system with minimum amount of compromised load. For example, the Class A attack destabilizes the power system while only compromising 2\% of the system loads. 
There were some cases in which the system did not converge, due to a limitation of the simulator.  However, using time domain simulation helps finding the oscillating frequencies that are more susceptible to this attack.   


\begin{figure}[t]
\includegraphics[width=0.5\textwidth]{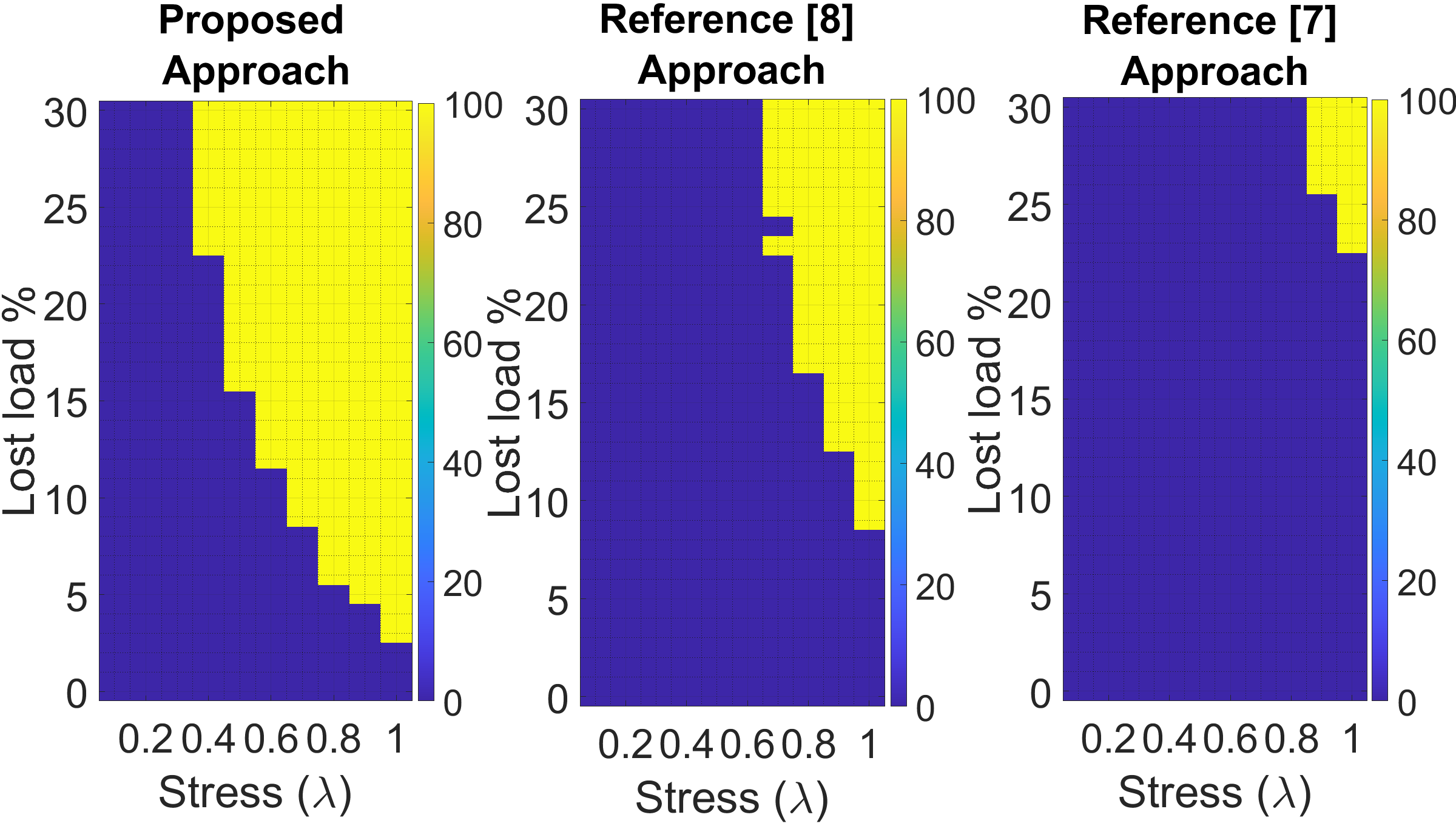}
\caption{The percentage of lost loads for load oscillation attacks with variable compromised load percentage and system stress level for the present approach compare to \cite{carter, Anu}. }
\label{fig: comp_all2}
\end{figure} 

\subsubsection{Validation} Following the same procedure as for the IEEE RTS 96 test system, we apply our proposed attack and the benchmarks to the Polish test system. In addition, the same metric of the percentage of lost load is used for comparison, as shown in Figure \ref{fig: comp_all2}. 
The results show 
that the proposed attack causes more damage than the benchmark at both high and low stress levels. Although the proposed attack and the benchmark \cite{carter} approach attack the same area, the proposed attack causes more damage using a smaller amount of compromised load. From Figure \ref{fig: comp_all2} we can see that the attack presented by \cite{carter} has no effect on the system when $\lambda\leq 0.7$, whereas our proposed attack is able to cause damage to the system as long as $\lambda\geq 0.4$. In addition, at $\lambda = 1$ the proposed attack succeeds in disturbing the power system by only compromising 3\% of the system load compared to 9\% for the attack in \cite{carter}. 

From Figure \ref{fig: comp_all2}, we can see that distributing the adversary capability throughout all system loads weakens the attack. We apply the attack presented  by \cite{Anu} on the Polish test system, failing to cause any damage when $\lambda\leq 0.09$. In addition, distributing the adversarial capability in a large system required compromising at least 25\% of the system load to cause a disturbance to the system.

\section{Conclusions} 
This paper demonstrates a load oscillation attack that causes the power system to lose a considerable amount of its load. 
Results of this work demonstrate that with a proper optimization of attack parameters, a load oscillation attack can be successful with much smaller oscillation magnitude than what had been reported as required in previous studies. We show that using fast and medium oscillating frequencies for the attack causes more damage than using a single oscillating attack frequency or fast and slow oscillating attack frequencies. Additionally, we illustrate the impact and characteristics of the attacks using the RTS-96 bus system and we verify them using the Polish 2383-Bus system. We show that the proposed attack succeeds in causing a major disturbance in the IEEE RTS-96 and Polish bus systems by oscillating 8\%  and 3\% respectively, of the system load. Lastly, we show that a traditional emergency control can not mitigate the damage caused by these types of attacks and further study of the mitigation strategies is recommended.

\appendix[System Loading (Stress Level)]

To create various levels of system stress for the RTS 96 and Polish 2383 bus systems, we multiply the active power components of the system loads ($\mathbf{d}_i $) and generators ($\mathbf{g}_k $) by a multiplier $\lambda$. Additionally, we factor each turbine governor's maximum output power ($\mathbf{tg}^{max} $) with ($\lambda \times 1.1$). The following equations, drawn from \cite{Anu}, describe the process of changing the system stress levels:

\begin{equation}
\mathbf{d}_i=\lambda \times \mathbf{d}_i^0, \forall ~i \in \mathcal{L}
\end{equation}
\begin{equation}
\mathbf{g}_k=\lambda \times \mathbf{g}_k^0, \forall ~k \in \mathcal{G}
\end{equation}          
\begin{equation}
\mathbf{tg}^{max}_k=\lambda \times~1.1~\times \mathbf{tg}^{0~max}_k, \forall ~k \in \mathcal{G}
\end{equation} 
where $\mathbf{d}_i^0,~ \mathbf{g}_k^0,~$and $\mathbf{tg}^{0~max}_k $ are the base case active power components of the $ i^{th}$ load, $k^{th}$ generator and the $k^{th}$ governor base case maximum active power output limit respectively. It is important to note that by scaling the generator output and turbine governor limits as we scale the load, we create a stress state without placing artificial strain on system controls. If the generation and turbine governor limits do not scale up with the load, then the system controls will handle the entire burden, which will create vulnerability in the system prior to any external disturbance. Allowing generation and governor limits to ramp up or down as the system loads increase or decrease, helps to isolate the effect of the oscillation attack on the power system. Additionally, each governor limit is scaled by $1.1 \times \lambda$ in order to capture the flexibility around the dispatch point.

\bibliographystyle{IEEEtran}
\bibliography{oscillatory}

\end{document}